# Heart Rate Variability Patterns reflect Yoga Intervention in Chronically Stressed Pregnant Women: A Quasi-Randomized Controlled Trial

**Marlene J. E. Mayer** [1†]**, Nicolas B. Garnier** [2†]**, Clara Becker** [3]**, Marta C. Antonelli** [4]**, Silvia M. Lobmaier** [5†,] **and Martin G. Frasch** [6†*]

[1] TUM university hospital, department of obstetrics and gynecology, Technical University of Munich, Germany, marlene.mayer@tum.de
[2] CNRS, ENS de Lyon, LPENSL, UMR5672, 69342, Lyon cedex 07, France, nicolas.garnier@ens-lyon.fr
[3] TUM university hospital, department of obstetrics and gynecology, Technical University of Munich, Germany, clara.becker@tum.de
[4] Instituto de Biología Celular y Neurociencia "Prof. Eduardo De Robertis", Facultad de Medicina, Universidad, de Buenos Aires, Argentina, mca@fmed.uba.ar
[5] TUM university hospital, department of obstetrics and gynecology, Technical University of Munich, Germany, silvia.lobmaier@tum.de
[6] *Department of Obstetrics and Gynecology and Institute on Human Development and Disability, University of Washington, Seattle, USA, mfrasch@uw.edu, phone: +1.206.539.6900, Postal address: Department of Obstetrics and Gynecology, University of Washington, 1959 NE Pacific St, Box 356460, Seattle, WA 98195,
*Corresponding author.
† These authors contributed equally to this work.

## Abstract

Prenatal maternal stress (PS) is a risk factor for adverse offspring neurodevelopment. Heart rate variability (HRV) complexity provides a non-invasive marker of maternal autonomic regulation and may be influenced by mind–body interventions such as Yoga. In this quasi-randomized controlled trial, 28 chronically stressed pregnant women were followed from the second trimester until birth: 14 participated in weekly Hatha Yoga with electrocardiogram (ECG) recordings, and 14 received standard obstetric care with monthly ECGs. Group allocation was based on availability, with participants unaware of their assignment at enrollment. HRV complexity was assessed first with Sample Entropy and Entropy Rate and then expanded to 94 HRV metrics spanning temporal, frequency, nonlinear, and information-theoretical domains. All metrics were covariate-adjusted (maternal age, BMI, gestational age), standardized, and analyzed using timepoint-specific principal component analysis (PCA). From this, a unified HRV index was derived. Analyses revealed that HRV metric relationships changed dynamically across pregnancy, with PCA loadings shifting from frequency toward complexity measures in late gestation. The mixed effects model identified a significant time x group interaction effect (p=0.041). These findings suggest a restructuring of HRV signal-analytical domains with advancing pregnancy attributable to Yoga and highlight the utility of advanced HRV analysis frameworks for future, larger trials.

## 1. Introduction

Stress has a well-known negative impact on physical and mental health [1]. Pregnant women are in an even more vulnerable position, not just because of the challenging changes happening to the body, especially the hormone system, and life situation [2], but also because the experienced stress can impact the developing fetus' lifelong health as well.

The global prevalence of stress in pregnant women is estimated at up to 26% [3]. Various adverse effects of maternal prenatal stress (PS) on the child's development have been demonstrated: PS in the form of stressful life events causes lower motor competence in the offspring during late childhood and adolescence. The effect was the strongest if the stressful life events appeared in late pregnancy [4]. PS has been associated with psychiatric abnormalities, such as behavioral disorders or depression [5, 6]. Continuous PS throughout the first and second half of pregnancy causes lower verbal cognitive skills in 2-year-old offspring of both sexes, as well as lower motor skills for the male offspring [7].

Numerous possible mediators link PS to altered child development. One of the most established mediators is cortisol, since it can pass the placental barrier. Furthermore, other mediators such as catecholamines, cytokines, reactive oxygen species, and serotonin/tryptophan are suspected to impact offspring's development directly or indirectly [8]. New results show that PS also affects the fetal iron homeostasis [9]. Moreover, PS can lead to epigenetic changes in the form of DNA methylation, indicating a possible impact on the endocrine, immune, and central nervous systems [10]. In this way, PS can lead to altered development of the child's hypothalamic-pituitary-adrenal (HPA) axis [11], which can cause behavioral and cardiovascular pathologies [12].

Another affected system is the autonomic nervous system (ANS). PS exerts adverse effects on the development of the fetal ANS, which are detectable, for example, in an impaired heart rate variability (HRV) of the offspring before and after birth [13, 14].

In a previous study, Lobmaier et al. (2020) identified a novel biomarker of fetal ANS activity: Transabdominal electrocardiograms (aECG) were recorded, and bivariate phase-rectified signal averaging was used to find couplings between maternal and fetal heart rates (mHR, fHR). For mHR decreases, the fHR of stressed mothers also decreased, whereas the fHR of non-stressed mothers remained stable. This responsiveness of the fHR to the mHR was coined the fetal stress index (FSI) [15].

In this study, we measured PS by assessing the activity of maternal ANS using Sample Entropy (SampEn), Entropy Rate and 92 other heart rate variability markers. SampEn and Entropy Rate are related biomarkers that show the variability and unpredictability of mHR, with higher entropy indicating a healthy and adaptable ANS and lower entropy indicating stress. In an animal study, fetal acidosis in sheep fetuses was related to lower entropy rate values [16]. Other human studies support these results: intrapartum fHR analyses showed lower Approximate Entropy and SampEn in acidotic fetuses [17] and a decrease of SampEn as an early indicator of neonatal sepsis [18]. In adults, it was shown that HRV decreases when individuals are exposed to stress [19]. Byun et al. (2019) compared entropy values in a group of patients diagnosed with Major depressive disorder to a control group and showed that entropy features are lower in the depression group. The difference became even more visible in the recovery phase after mental stress induction. During a relaxation task, the entropy increased, indicating an enhanced vagal activity [20].

Building on our earlier publications, we apply the HRV code concept in the present study [21, 22].

Briefly, the analysis of HRV has evolved beyond traditional statistical measures to encompass a more sophisticated framework that views HRV as a potential complex biomarker of the brain-body communication. Recent theoretical advances propose the existence of an "HRV code" — a systematic encoding of physiological information within beat-to-beat cardiac intervals that reflects brain-body communication pathways [21-23]. This conceptual framework suggests that HRV contains signatures of information flow between organs and specific responses to physiological stimuli, exhibiting features of time structure, phase space organization, target specificity, and species universality. The HRV code paradigm posits that process-specific subsets of HRV measures indirectly map phase space traversals, reflecting the informational content required for physiological regulation. This perspective transforms HRV analysis from descriptive statistics to a bioinformatics approach for decoding the complex spatiotemporal dynamics of autonomic regulation, with particular relevance for understanding how interventions like Yoga may systematically alter the informational architecture of cardiac variability.

In the present study, we focused on the mHR dynamics because the underlying wearable electrocardiogram (ECG) technology is widely available and received health regulatory clearance in both the USA and the EU. Recent surveys in the USA indicate high acceptance of wearable ECG devices in pregnancy, with over 90 % of women willing to use them for maternal and fetal monitoring, highlighting their potential for clinical adoption [24].

Yoga and meditation can lower self-reported stress [25] and anxiety [26]. Yoga lowers physiological stress biomarkers such as cortisol, systolic blood pressure, and resting mHR, which improves the regulation of the ANS and HPA axis [27]. Meta-analyses suggested that Yoga, as a non-pharmacological intervention in pregnancy, can be beneficial for lowering stress, anxiety, and depression as well as improving the birth outcome with shorter duration of labor, less need for pain medication, and higher rate of vaginal births. Even more importantly, there was no evidence of adverse events in any trial, making it a safe practice for pregnancy [28, 29].

Therefore, in the present study, we used a Yoga intervention program to test whether such practice can reduce PS, assessed by maternal HRV biomarkers, and thereby diminish the above-described adverse effects on the fetus as well as on the mother. In concrete terms, our question was whether delta (i.e., change) analyses between the beginning and the end of the study show increased entropy biomarkers in the Yoga intervention group compared to the control group and whether broader patterns across the full set of 94 HRV metrics — including entropy measures — indicated changes in autonomic regulation.

## 2. Materials and Methods

*2.1. Design and Participants*

The quasi-randomized controlled trial was performed from December 2022 to May 2024 in Munich, Germany, at the University Hospital of the Technical University of Munich (TUM). For recruitment (Figure 1), we advertised our study on the homepage of TUM University Hospital, as well as in the referring physicians' centers and the Klinikum Dritter Orden. We screened 82 pregnant women with Cohen's Perceived Stress Scale (PSS) at 17 (control) vs 16 (Yoga) weeks of gestation.

To generate a cohort with stressed women, we set the PSS cut-off score at ≥19 as established [15]. The inclusion criteria were singleton pregnant women aged 18-45 years, 12-20 weeks of gestational age, fluent in German, and willing to participate in at least 75 % of the Yoga classes. The exclusion criteria included fetal malformations or serious placental alterations defined as fetal growth restriction [30], severe maternal illnesses [31], maternal drug or alcohol abuse, and preterm birth (less than 36 weeks). This yielded 28 study participants. Due to organizational constraints and the limited number of participants, true randomization was not feasible. Instead, participants were quasi-randomly allocated to either the Yoga or control group based on group availability at the time of enrollment. Participants were not informed of their group assignment before inclusion. Blinding of investigators was not possible, as they were responsible for planning and conducting the study. This approach resulted in 14 participants in the Yoga intervention group and 14 in the control group. These participants were followed between 21 and 35 weeks of gestation (control) vs 18 and 35 weeks of gestation (Yoga).

The control group received standard obstetric care. Yoga group participants attended a weekly 75-minute Yoga class with elements of Hatha Yoga and a weekly 20-minute online Yoga Nidra class with relaxation and mindfulness practice. During the Yoga class and an additional 15 minutes of lying still before and after the class, we recorded an aECG. Such placement of the ECG was chosen to conform to the standard maternal experience of fHR recording during a non-stress test. The ECG measurement was also performed once a month in the control group for 45 minutes. The procedure was performed at a similar time as the Yoga class, in the afternoon, to prevent confounding by the circadian rhythmicity of fetal behavioral states [32] and cortisol levels [33]. The recording duration in the control group was shorter than in the Yoga group to ensure participant compliance and maintain the feasibility of the recordings.

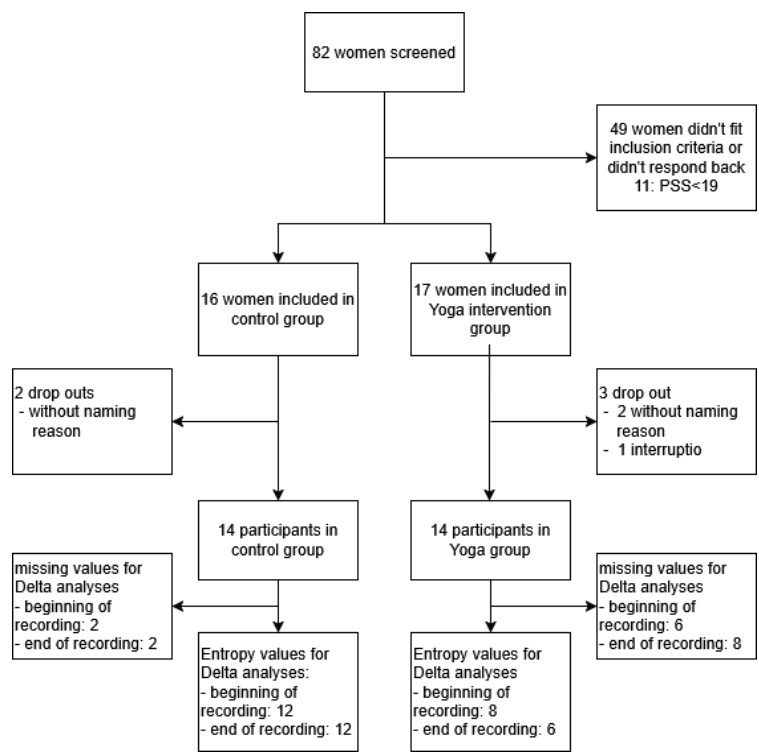

**Figure 1.** Recruitment flow chart.

*2.2. Questionnaires: Maternal Perceived Stress, Sociodemographic Questionnaire, Physical Activity Questionnaire*

PSS is a validated and well-established self-report tool to examine the nonspecific perception of stress [34]. The likewise validated German version consists of 10 questions assessing how overwhelming and uncontrollable life is perceived [35]. Therefore, each question must be answered on a 5-point scale, corresponding to 0 – 4 points. This adds up to an overall test result between 0 and 40 points, a higher number representing a higher perceived stress. No formal cut-off score exists for an individual being "stressed", as PSS is not a diagnostic tool.

For our study, we set the cut-off at 19 and higher, based on Lobmaier et al.'s (2020) previous study showing that 19 is the 80th % percentile of the PSS-10 [15].

Furthermore, we evaluated sociodemographic data, including household income, education, marital status, and employment status.

For the second-trimester screening (18+0 to 21+6 weeks of gestation) and the third-trimester screening (28+0 to 31+6 weeks), we evaluated physical activity and Yoga practice both within and outside the study. By these time points, most participants in the Yoga group had already been enrolled in the study.

### 2.3. Hatha Yoga and Yoga Nidra

The Hatha Yoga class took place once a week in the afternoon as an "in-person class". Each class started with 15 minutes of lying still, followed by another 15 minutes of lying still accompanied by meditation and breathing exercises (Pranayama), 45 minutes of active Yoga, and ended with another 15 minutes of lying meditation and 15 minutes of lying still (Shavasana), resulting in a total recording time of 105 min. Each lesson covered different lectures and intentions in the meditation but contained the same Asanas in some variation. A certified prenatal Yoga teacher created the content to fit the needs of pregnant women. The active Yoga lessons started with grounding and opening by seated postures, tabletop position, forearm position, child's pose, and mobilization of the entire body. It was continued by modified sun salutations, standing postures, gentle back-and-forward bends, and an upside-down pose.

The Yoga Nidra class also took place once a week as a 20-minute online class. Here, we intended to lower the stress by integrating more mindfulness and relaxation into the daily lives of pregnant women. Yoga Nidra is practiced lying still on the back or one side. It started with breathing exercises, followed by a personal resolution/purpose for the lesson, a body scan, imagination exercises, and, in the end, a repetition of the resolution and return to the environment [36].

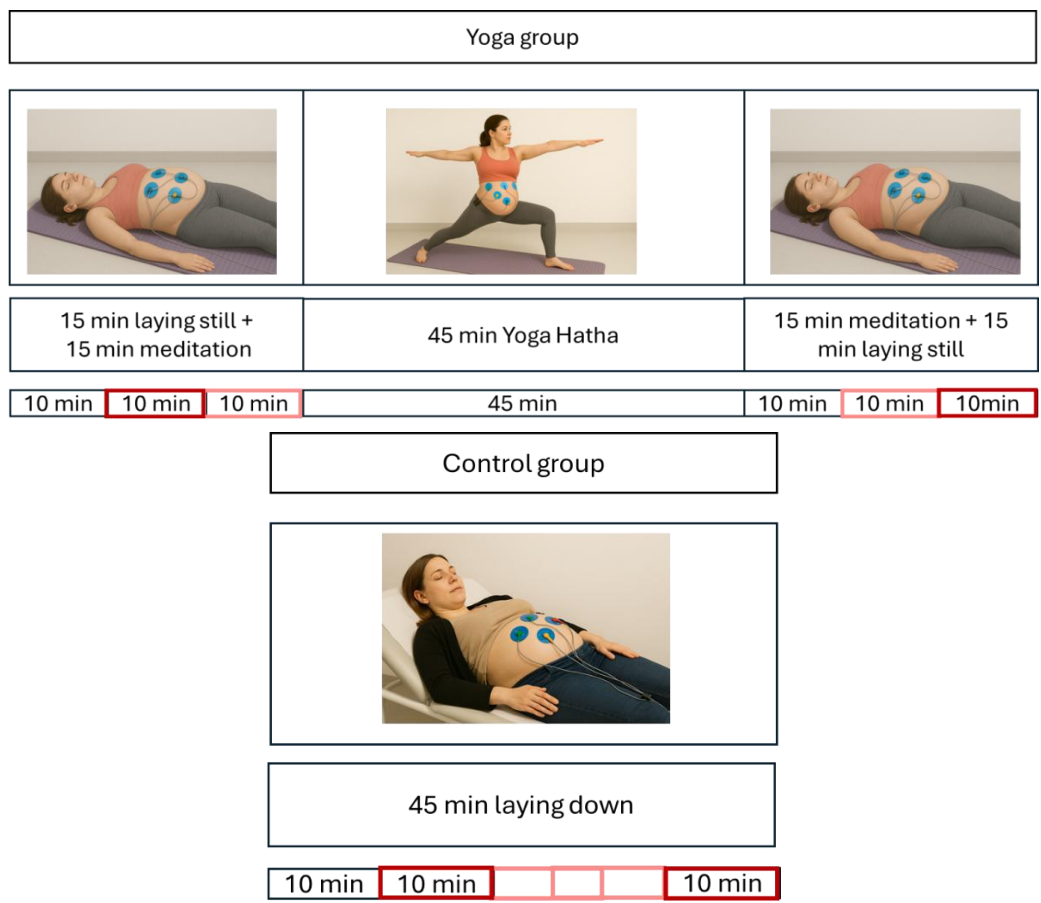

**Figure 2.** Electrocardiogram's (ECG) time windows of an individual experiment chosen for the entropy analyses. Dark red is the preferred time window; light red is the alternative time window when the signal quality of the other window was not sufficient. Pictures are AI-generated. Our goal was to choose baseline or the latest possible recovery segments to avoid signal contamination by exercise.

### 2.4. ECG acquisition and SampEn and Entropy Rate computation

We used wearable, clinical-grade ECG devices (Bittium Faros 360) to acquire the maternal abdominal ECG at 1000 Hz sampling rate. These devices give an instantaneous measure of the RR peak intervals from which we computed the heart rate (HR), which we then averaged over time-windows of size τ=2s and resampled at a frequency of 20 Hz.

To obtain insight into the information content of the HR signals, we measured their SampEn [18, 37], as well as their Entropy Rate [16, 17, 38] over the time-scale τ. Both SampEn and Entropy Rate quantify the HR variability. The Entropy Rate h(τ) is defined as the difference

$$h(\tau) = H(X(t), X(t-\tau)) - H(X(t-\tau)) = H(X(t) \mid X(t-\tau))$$

between the Shannon Entropy H(X(t)) of the low-pass filtered HR signal X(t) and the Joint Entropy H (X(t), X(t−τ)) of the same HR signal considered at two successive times t−τ and t separated by the time-scale τ = 2s.

SampEn and Entropy Rate measure the information content of the HR signal by quantifying its uncertainty or, equivalently, its complexity: they measure how much unknown information (in Shannon's meaning) is present in X(t) that was not present in X(t-τ). We assume here that HR can take any real value, and hence, we use continuous entropy. Consequently, values of the entropy — and thus of the Entropy Rate — depend on the relative magnitude of the signal fluctuations and can be negative. Albeit similar to Approximate Entropy and SampEn (which assume discrete data like traditional Shannon Entropy Rate and are hence always positive), the Entropy Rate is a superior estimate for the HR complexity to detect anomalies [17, 38]. Its statistical properties also make it more robust to missing data points [16].

Entropy can be understood as a measure of variability or diversity within a system. Low entropy indicates uniformity and predictability (measurements are very similar), while high entropy reflects heterogeneity and multiple possible outcomes [39]. Conceptually, this idea can be applied to physiological signals such as ECGs or HRV, where higher entropy indicates more complex or less predictable patterns, whereas lower entropy reflects more regular and stable signals.

SampEn and Entropy Rate were computed on HR time series in time windows of 10 minutes in non-overlapping windows. Each recording received a signal confidence index between 0 and 1, 1 being the most trustworthy. Cases with a confidence index lower than 0.9 were excluded from the analyses. The choice of the cut-off at 0.9 ensures that the bias in the entropy rate computation (as well as in the SampEn computation) is negligible (typically less than 0.02) [40].

The beginning of the recording reflects minutes 10 to 20 of the recording (unless it was replaced by minutes 20-30 due to a low confidence index). The end of the recording reflects the last 10 minutes of the recording (unless it was replaced by the penultimate 10 minutes due to a low confidence index). See Figure 2.

We computed the "delta" values as representing the temporal changes in the individual entropy metrics' values as follows. Delta was defined as a change in entropy value (SampEn or Entropy Rate) between the respective begin or end steady-state measurement at the end of the study (i.e., at 35 weeks of gestation) and at the start of the study (i.e., at 21 (control) vs 18 (Yoga) weeks of gestation). See Supplementary Figure S1.

*2.5. Heart Rate Variability (HRV) Computation Approach*

A total of 94 HRV metrics (See Supplementary Table S1) were derived from the R–R interval time series, spanning four physiological domains:
1. Temporal Domain (25 metrics): Traditional statistical measures of R-R interval variability, including Mean NN, SDNN, RMSSD, pNN50, pNN20, SDANN (1-, 2-, and 5-minute segments), coefficient of variation measures, percentile-based indices, and geometric measures (HTI, TINN).
2. Frequency Domain (6 metrics): Power frequency analysis yielding low frequency (LF, 0.04-0.15 Hz), high frequency (HF, 0.15-0.4 Hz), total power (TP, 0-0.4 Hz), LF/HF ratio, and normalized units (LFnu, HFnu).
3. Complexity/Information Domain (54 metrics): Nonlinear dynamics measures including Poincaré plot indices (SD1, SD2), entropy measures (approximate entropy [ApEn], sample entropy [SampEn], Shannon entropy, fuzzy entropy, multiscale entropy variants), detrended fluctuation analysis (DFA) scaling exponents ($\alpha1$, $\alpha2$), multifractal DFA parameters, correlation dimensions, fractal dimensions (Higuchi, Katz), complexity indices (CSI, CVI), and Lempel-Ziv complexity. Entropy rate measures the information generation rate (for details, see the respective subsection).

4. Specialized Domain (9 metrics): Additional cardiac measures including heart rate turbulence parameters, coefficient of variation, temporal variability indices, and frequency characteristics (centroid frequency, bandwidth).

The methodological details of this pipeline have been reported elsewhere, along with the reproducible code [41].

All raw HRV metrics were carefully quality-controlled, including systematic detection and appropriate treatment of missing values and outliers.

To ensure that our analysis captured the true intervention effect rather than the influence of confounding variables, we performed a two-step adjustment process.

First, each of the HRV metrics was individually adjusted for the effects of maternal age, gestational age, and BMI. This was achieved by fitting a linear regression model for each metric with the confounders as predictors. The resulting residuals, which represent the portion of HRV variance unexplained by these covariates, were carried forward for analysis.

$$\text{Model: HRV\_metric} = \beta_0 + \beta_1(\text{age}) + \beta_2(\text{gestational age}) + \beta_3(\text{BMI}) + \varepsilon$$

In the second step, the residualized HRV metrics were then transformed into z-scores (mean of 0, standard deviation of 1). This standardization step ensures that all metrics contribute equally to the subsequent Principal Component Analysis (PCA), regardless of their original scale. We validated this adjustment by confirming that the final residualized metrics showed minimal correlation ($|r| < 0.1$) with the confounders, while preserving the original intercorrelation structure among the HRV metrics themselves.

Recognizing that HRV is a dynamic process, we rejected a "one-size-fits-all" approach. Instead, we performed a separate PCA for each of the four distinct measurement timepoints (first recording- begin, first recording- end, last recording- begin, last recording- end). For each time point, PCA was performed on the full set of adjusted and standardized HRV metrics. Principal components (PCs) were selected based on their ability to explain a cumulative variance of at least 80%. This data-driven approach allows for the number of selected PCs to vary between timepoints, reflecting the changing structure of HRV dynamics. For each time point, a unified_HRV_index was constructed by summing the scores of the selected PCs. This created a single composite score per subject for each of the four timepoints, representing their overall autonomic complexity, independent of confounders.

To further justify our timepoint-specific PCA strategy, we examined the inter-metric correlation structure separately at each of the four measurement timepoints (first visit beginning, first visit end, last visit beginning, last visit end; see Figure 4 A-B). For each timepoint, we generated a correlation matrix using the adjusted HRV metrics. The overall correlation structure was quantified by calculating mean absolute correlations and identifying pairs of metrics with high collinearity ($|r| \geq 0.8$). In addition, we assessed changes in correlation structure between timepoints to evaluate potential network simplification (or vice versa) during pregnancy progression or as an effect of Yoga treatment.

These analyses revealed that the patterns of collinearity among HRV metrics were not stable but shifted significantly across the four timepoints. For example, metrics that were highly correlated at the beginning of the first visit showed weaker correlations by the end of the last visit. This evidence of dynamic collinearity strongly supports the use of timepoint-specific PCA, as a single, universal PCA structure would have obscured these important temporal shifts in autonomic regulation.

To complement these analyses, HRV metrics were further grouped by physiological domains (temporal, frequency, complexity/information, and specialized). Domain-specific contributions to the principal components were quantified by calculating mean absolute loadings within each domain. Cross-timepoint stability was then assessed by comparing domain contributions between the first and last visits, with percentage changes expressed as relative differences.

## 2.6. Statistical Methods

We first conducted the conventional analyses as prespecified in the study protocol. We used the Shapiro-Wilk test to test the continuous data for normal distribution.

Since our hypothesis assumes a directional effect of Yoga on HRV metrics (SampEn and Entropy Rate) as ANS biomarkers, we used one-sided statistical tests. For normally distributed data, we used the t-test; for not normally distributed data, we used the Mann-Whitney U test for group comparison. Categorical data were analyzed using Fisher's exact test (one- or two-tailed, depending on the hypothesis) for 2×2 contingency tables, and the chi-squared test for all other contingency tables. Group differences with p<0.05 were considered statistically significant and are marked in bold. Results are reported as mean (SD) or median (IQR), as appropriate. Missing data were handled by listwise deletion. These tests were conducted using IBM SPSS Statistics Version 29.0.2.0.

The study consisted of two primary endpoints: the PSS scores (the results are discussed in a different paper) and the entropy values. The sample size was determined a priori. Based on Satyapriya et al. (2009), we assumed a mean difference of 5 points in the PSS score with a within-group SD of 6 points, corresponding to Cohen's d=0.83 [42]. Because the two primary outcomes were predefined, a Bonferroni correction was applied, and the per-outcome significance level was set to $\alpha$ = 0.025. To obtain 80% power, the required sample size was calculated as 28 participants per group. Due to recruitment constraints, only 28 participants in total (14 per group) were enrolled. Using a significance level of alpha = 5 %, this leads to a power of 56%.

To further investigate the intervention effect on HRV while accounting for repeated measures and potential confounders, we additionally applied linear mixed-effects models separately for Sample Entropy and Entropy Rate. Each model included fixed effects for group (Yoga vs. Control), time (first vs. last measurement), and the Group × Time interaction as the primary test of the intervention effect. A random intercept for each participant accounted for individual baseline differences and the repeated-measures design. Sensitivity analyses included maternal age, BMI, and gestational age as covariates.

For the comprehensive HRV analysis, we computed 94 HRV metrics across temporal, frequency, complexity/information-theoretic, and specialized domains. After covariate adjustment and timepoint-specific principal component analysis, using PCA we derived a unified HRV index at four measurement periods (first visit–beginning, first visit–end, last visit–beginning, last visit–end).

Due to the small sample size and to test the putative Yoga intervention effect most directly, we focused on the two key timepoints with maximal interpretability (beginning of the first visit and end of the last visit) and fitted a linear mixed-effects model specified as:

unified_hrv_index ~ group × timepoint + (1 | subject_id)

This specification explicitly tests the group × timepoint interaction, which would indicate whether the Yoga intervention altered the trajectory of HRV complexity over time compared to the control group. These extended analyses were conducted using Python 3.x (pandas, numpy, scipy, scikit-learn, statsmodels). Statistical significance was set at $\alpha$ = 0.05 for all tests. The analysis pipeline and code are openly available at https://github.com/martinfrasch/hrv

In line with recent statistical recommendations, in the final analyses, we did not apply strict significance thresholds or adjustments for multiple comparisons in this exploratory study. As emphasized by Amrhein et al. (2019), dividing into "statistically significant" versus "non-significant" based on an arbitrary p-value cut-off (e.g., p < 0.05) is misleading and can obscure meaningful patterns in the data [43]. Instead, the focus should be placed on effect sizes, confidence intervals, and the broader evidential context rather than on binary significance testing. Especially as our study has a very small sample size, the primary aim has to be hypothesis generation rather than definitive validation.

## 3. Results

### 3.1. Group Characteristics

Table 1. Group Characteristics.

| p-value | Yoga<br>N=14 | Control<br>N=14 | |
|---|---|---|---|
| 0.284 | 15.9 (3.0) | 17.3 (3.4) | Gestational age at inclusion (weeks), mean (SD) |
| 0.283 | 18.3 (3.5) | 21.0 (16.6-21.9) | Gestational age at first ECG (weeks), median (IQR)/ mean (SD) |
| 0.612 | 34.6 (3.6) | 35.3 (2.5) | Gestational age at last ECG (weeks), mean (SD) |
| 0.157 | 33.3 (3.1) | 35.2 (3.8) | Maternal age at inclusion (years), mean (SD) |
| 0.828 | 22.5 (20-28) | 24.86 (4.8) | PSS at inclusion, mean (SD), median (IQR) |
| 0.867 | 21.9 (3.4) | 21.9 (2.5) | BMI at inclusion [1], mean (SD) |
| 0.219 | 12 (86) | 13 (93) | Ethnicity European, n (%) |
| 0.420 | | | Marital status, n (%) |
| | 8 (57) | 11 (79) |   Married |
| | 6 (43) | 3 (21) |   In relationship |
| | 0 (0) | 0 (0) | Single mom, n (%) |
| 0.648 | 12 (86) | 10 (71) | Working, n (%) |
| 0.097 | 14 (100) | 10 (71) | Highest level of education: University degree, n (%) |
| 0.675 | | | Net household income, n (%) |
| | 1 (7) | 0 (0) |   1000-2500€ |
| | 5 (36) | 5 (36) |   2500-5000€ |
| | 5 (36) | 7 (50) |   5000-10000€ |
| | 3 (21) | 2 (14) |   >10000€ |
| 0.165 | | | Parity, n (%) |
| | 13 (93) | 9 (64) |   Primipara |
| | 1 (7) | 5 (36) |   Multipara |
| 1.00 | 12 (86) | 12 (86) | Planned pregnancy, n (%) |
| | 0 (0) | 0 (0) | Substance use (alcohol, tobacco, or drugs), n (%) |
| 1.000 | 1 (7) | 2 (14) | Autoimmune disease, n (%) |
| | 0 | 0 | Gestational diabetes, n (%) |
| 1.0 | 0 | 1 (7) | Arterial hypertension, n (%) |
| 0.481 | 14 (100) | 12 (86) | Physical activity before pregnancy, n (%) |
| 1.000 | 10 (71) | 10 (71) | Yoga experience before study, n (%) |
| **0.005** | 12 (100) | 7 (50) | Physical activity at second-trimester screening (incl. Yoga)[2], n (%) |
| 0.241 | 14 (100) | 12 (86) | Physical activity at third-trimester screening (incl. Yoga), n (%) |
| **0.008** | 14 (100) | 8 (57) | Third trimester: Yoga practice (private or study-related), n (%) |
| **< 0.001** | | | Frequency of Yoga practice during third trimester, n (%) |
| | 0 | 6 (43) |   0 min |
| | 0 | 3 (21) |   1-30 min |
| | 0 | 3 (21) |   31-60 min |
| | 12 (86) | 2 (14) |   61-90 min |
| | 1 (7) | 0 |   91-120 min |
| | 1 (7) | 0 |   181-210 min |
| 0.706 | 8 (57) | 6 (43) | Sex of newborn: female, n (%) |

Continuous variables are presented as mean (SD) if normally distributed or as median (IQR) if skewed; categorical variables are presented as n (%).

[1] missing value of 1 Yoga group participant

[2] missing values of 2 Yoga group participants

There were no significant differences in the maternal characteristics. Our study population was a highly educated cohort with an above-average income and high sports and Yoga activity levels in both groups. We found the physical activity at the second trimester screening timepoint, as well as the Yoga practice and frequency in 3rd trimester, to be higher in the Yoga group compared to the control group (all p <0.05).

*3.2. Data quality*

The ECG device did not always provide the heart rate time series along with the raw ECG. Since we relied on the heart rate time series for our analyses, this resulted in two subjects being removed in the control (e.g., the first or the last session heart rate data missing, so the delta cannot be computed, so we have 12 controls) group and four subjects missing in the Yoga group (resulting in ten subjects in this group). In addition, the SQI threshold of 0.9 resulted in removing four additional subjects in the Yoga group. For the control group, 3 windows out of 56 were excluded and were replaced by the previous (for the "last") or the next (for the "first"), so that the number of subjects did not change. See Table 2.

|  |  | total | valid begin | valid end |  |
|---|---|---|---|---|---|
| Yoga | first session | 10 | 10 | 10 | |
| Yoga | last session | 10 | 8 | 6 | |
| Yoga | delta | 10 | 8 | 6 | ← delta needs both first and last |
| control | first session | 12 | 12 | 12 | |
| control | last session | 12 | 12 | 12 | |
| control | delta | 12 | 12 | 12 | |

**Table 2.** Usable data for the ECG recordings. Note the drop of sample size for delta at the final measurement due the requirement of needing the matching first and last recording.

*3.3. Primary Outcomes*

*3.3.1. Sample Entropy and Entropy Rate*

The mean participation rate at Hatha Yoga classes was 76% (12%). The gestational week at the first ECG recording at the study start was similar in both groups, averaging 20. At the last visit, the gestational week was also similar at 35. SampEn and Entropy Rates were higher in the Yoga group at the first and the last visits. See Table 3.

**Table 3.** Sample Entropy and Entropy Rate of maternal heart rate for the first and last visit of the study during resting conditions.

**Sample Entropy**
**First visit**

|  | Yoga | Control | p-value | Cohen's d | Effect size r |
|---|---|---|---|---|---|
| Begin | n=10 | n=12 | **0.011** | 1.19 | 0.51 |
|  | 3.03 (0.43) | 2.55 (0.38) |  |  |  |
| End | n=10 | n=12 | **0.006** | 1.19 | 0.51 |
|  | 3.14 (0.59) | 2.59 (0.32) |  |  |  |

**Last visit**

|  | Yoga | Control | p-value | Cohen's d | Effect size r |
|---|---|---|---|---|---|
| Begin | n=8 | n=12 | **<0.001** | 1.69 | 0.64 |
|  | 3.19 (0.51) | 2.50 (0.33) |  |  |  |
| End | n=6 | n=12 | **0.030** | 1.43 | 0.58 |
|  | 3.09 (0.56) | 2.54 (0.27) |  |  |  |

**Entropy Rate**
**First visit**

|  | Yoga | Control | p-value | Cohen's d | Effect size r |
|---|---|---|---|---|---|
| Begin | n=10 | n=12 | **0.004** | 1.37 | 0.57 |
|  | 2.45 (0.49) | 1.85 (0.39) |  |  |  |
| End | n=10 | n=12 | **<0.001** | 1.57 | 0.62 |
|  | 2.58 (0.53) | 1.88 (0.36) |  |  |  |

**Last visit**

|  | Yoga | Control | p-value | Cohen's d | Effect size r |
|---|---|---|---|---|---|
| Begin | n=8<br>2.98 (0.58) | n=12<br>1.90 (0.36) | **<0.001** | 2.36 | 0.76 |
| End | n=6<br>2.68 (0.35) | n=12<br>1.85 (1.7-2.1) | **0.003** |  | 0.66 |

Continuous variables are presented as mean (SD) if normally distributed or as median (IQR) if skewed. For independent t-tests, effect sizes are reported as Cohen's d (with corresponding effect size r additionally shown for comparability). For Mann–Whitney U tests, effect sizes are reported as r, because Cohen's d is not appropriate for rank-based tests.

In the delta analyses (Table 4), we considered the change in entropy values between the beginning and the end of the study, comparing the groups. When considering the beginning of the recordings, the SampEn did not change significantly; the Entropy Rate increased for the Yoga group (p=0.034). At the end of the recording, we saw no significant differences for either SampEn or the Entropy Rate.

In summary, changes were more noticeable at the beginning of the recordings than at the end. See Table 4 and Figure 3.

**Table 4.** Delta analyses compare the entropy between the begin and the end of the study.

**Sample Entropy**

|  | Yoga | Control | p-value | Cohen's d | Effect size r |
|---|---|---|---|---|---|
| Begin | n=8<br>0.18 (0.32) | n=12<br>-0.53 (0.34) | 0.072 | 2.14 | 0,73 |
| End | n=6<br>0.08 (0.36) | n=12<br>-0.05 (0.18) | 0.159 | 0.52 | 0.25 |

**Entropy Rate**

|  | Yoga | Control | p-value | Cohen's d | Effect size r |
|---|---|---|---|---|---|
| Begin | n=8<br>0.50 (0.67) | n=12<br>0.05 (0.39) | **0.034** | 0.87 | 0,40 |
| End | n=6<br>0.19 (0.57) | n=12<br>0.11 (0.34) | 0.360 | 0.19 | 0.094 |

Continuous variables are presented as mean (SD) if normally distributed or as median (IQR) if skewed. For independent t-tests, effect sizes are reported as Cohen's d (with corresponding effect size r additionally shown for comparability).

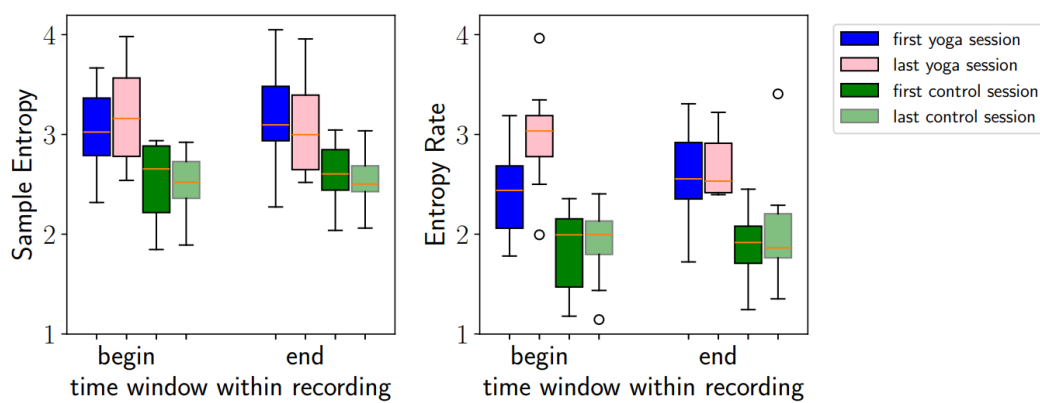

**Figure 3.** Sample Entropy (SampEn) and Entropy Rate in the second ("begin") and last ("end") 10-minute time windows are compared between the groups (Yoga vs. control) and during pregnancy. Blue (pink): first (last) session for the Yoga group; green (light green): first (last) session for the control group. At the first session, both entropy biomarkers were already higher in the Yoga group compared to the control group. By the end of the study, the Entropy Rate had further increased in the Yoga group, while it remained stable in the control group.

In a subsequent step, we re-analyzed the intervention effect on SampEn and Entropy Rate using linear mixed-effects models adjusted for maternal age, BMI, and gestational age. Under this more robust framework, the previously observed effects were no longer significant.
(Group × Time interaction: SampEn β = 0.069, p = 0.537; Entropy Rate β = 0.098, p = 0.401). Instead, the models confirmed substantial baseline differences between groups (SampEn p = 0.009; Entropy Rate p = 0.001), indicating that the apparent effects were primarily driven by pre-existing allocation bias rather than the intervention itself. In addition, gestational age emerged as a significant predictor of Sample Entropy (p = 0.038).

### 3.3.2. Comprehensive HRV analysis

Correlation analyses across the four measurement timepoints (first recording beginning, first recording end, last recording beginning, last recording end) demonstrated that HRV metrics relationships were dynamic rather than static. Mean absolute correlation coefficients ranged from 0.364 to 0.426 across these timepoints, with significant differences detected in 4 of 6 pairwise comparisons (p < 0.05). Also, we observed a 36% reduction in high correlations during pregnancy progression (454 → 290 pairs). Between 8.4% and 10.7% of metric pairs exhibited high collinearity ($|r| \geq 0.8$), indicating considerable but variable redundancy in the HRV metrics space. See Figure 5 A-B.

To account for this temporal variability, we performed timepoint-specific principal component analyses. These analyses confirmed that the structure of HRV variability changed across measurement periods. To reach the 80% variance threshold, the number of principal components required ranged from three at the beginning and end of the first recording to six at the beginning of the last recording. As expected, the first principal component explained the largest share of variance (31.2–38.7%), yet this proportion was insufficient to capture the majority of HRV variability on its own. This finding underscores that HRV is inherently multidimensional and that meaningful indices must integrate several components rather than relying on a single factor. See Figure 4 C-D. Domain-specific analyses further revealed that frequency measures became markedly less influential (−81.2%) from early to late pregnancy, while complexity measures gained importance (+32.8%), and time-domain metrics remained stable (−1.6%). This restructuring highlights pregnancy-associated shifts in autonomic regulation (Figure 4E).

Based on the PCA results, we derived a unified HRV index by summing the principal components that together explained at least 80% of the variance at each measurement time point. This index condenses the multidimensional HRV structure into a single score per participant and timepoint, while preserving temporal specificity. By accounting for dynamic collinearity, the unified index provided a robust framework for comparing HRV between groups and addressed a key limitation of conventional longitudinal HRV analyses, which typically assume static correlation structures.

We next applied the statistical framework to the unified HRV index. Within-group analyses showed no significant change from beginning to the end of the study (See Figure 4 F-G). Longitudinal analysis using mixed-effects modeling confirmed a significant time × group interaction (β = −6.817, p = 0.041) alongside a strong baseline group effect (β = 12.855, p = 0.001), indicating that Yoga participation influenced the trajectory of HRV across pregnancy (Figure 4H).

To place these results in context, we highlight five main findings emerging from our advanced methodological framework:

1. Correlation network simplification: a 36% reduction in high correlations during pregnancy progression (454 → 290 pairs).

2. HRV domain restructuring: frequency measures decreased (−81.2%) while complexity measures increased (+32.8%) in late pregnancy.

3. Baseline differences: significant initial difference of the Yoga group in the unified HRV index (p = 0.001).

4. Differential trajectories: significant time × group interaction (p = 0.041), suggesting that Yoga modulated the pattern of HRV changes.

5. Effect sizes: moderate for Yoga group changes (d = −0.521) versus small for controls (d = 0.106).

Together, these results suggest that prenatal Yoga practice may influence autonomic cardiovascular regulation, with evidence for distinct trajectories in HRV between intervention and control groups.

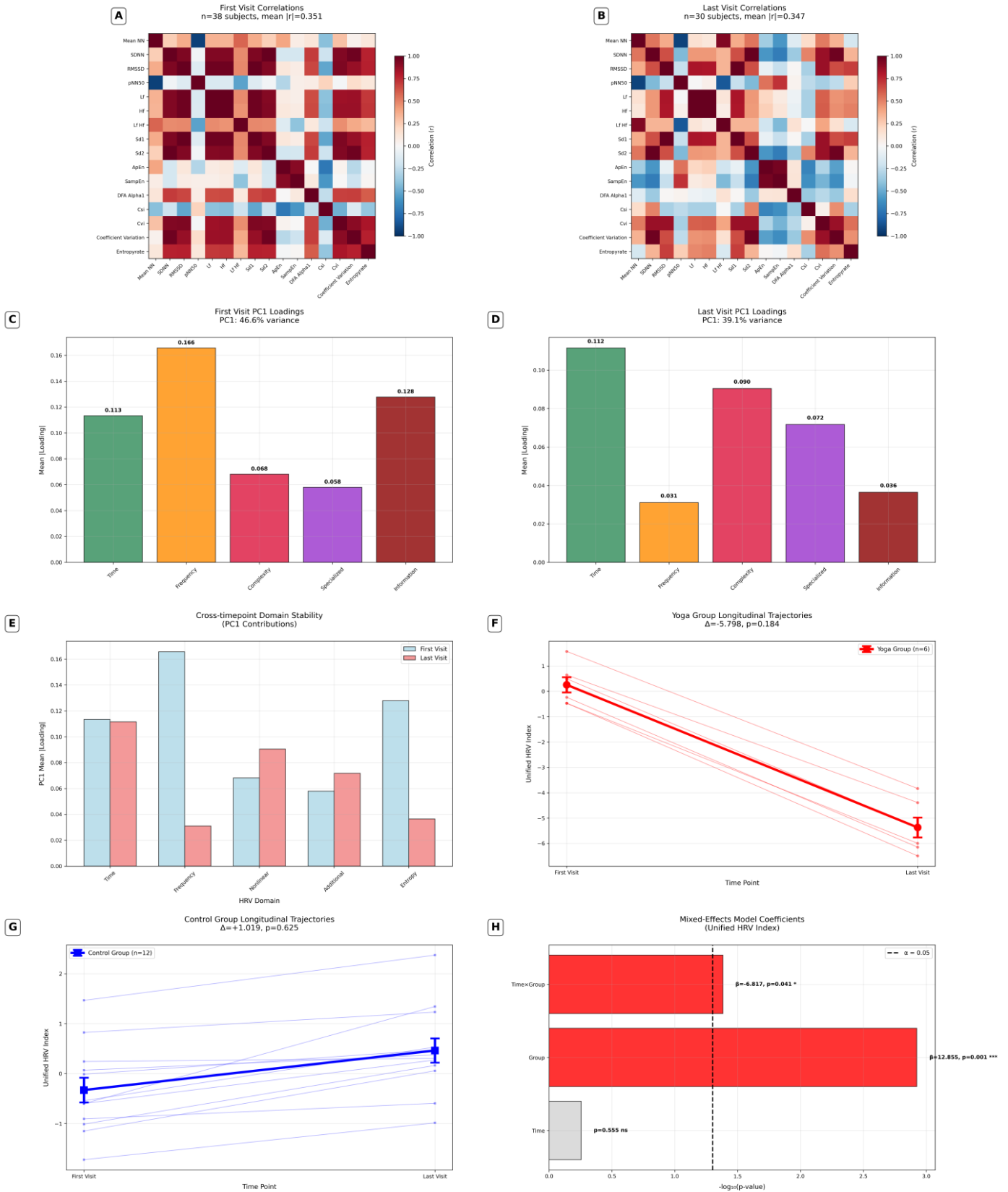

**Figure 4.** Comprehensive HRV analysis in pregnancy: correlation structure, PCA, domain contributions, and longitudinal modeling.

(A–B) Correlation matrices of representative HRV metrics at the first visit (A) and last visit (B). Colors indicate correlation coefficients from −1 (blue) to +1 (red).

(C–D) Principal component analysis (PCA) loadings for the first principal component (PC1) at the first (C) and last (D) visits. HRV metrics are grouped by physiological domains: Time (green), Frequency (orange), Complexity (red), Specialized (purple), and Information-theoretic (dark red).

(E) Cross-timepoint comparison of domain contributions to PC1 between the first and last visits.

(F–G) Distribution and longitudinal trajectories of the unified HRV index for the Yoga group (F) and Control group (G).

(H) Results of linear mixed-effects models on the unified HRV index, showing standardized coefficients for group, time, and interaction terms.

## 4. Discussion

### 4.1. Key Findings and Interpretation

#### 4.1.1. SampEn and Entropy Rate

We detected a group difference (p= 0.034) for the Entropy Rate delta analyses at the beginning of the recording, i.e., the baseline recording prior to Yoga intervention at 20 and 35 weeks of gestation. The Entropy Rate increased in the Yoga group, while it remained stable in the control group. These findings may indicate a potential effect of Yoga on maternal ANS.

When using linear mixed-effects models adjusting for maternal age, BMI, and gestational age, previously observed differences in Sample Entropy and Entropy Rate were no longer significant. These findings suggest that above individual results are largely explained by baseline group differences in these HRV metrics rather than by the intervention itself. Larger cohorts are required to provide more definitive insights.

Interestingly, in the individual tests, the effect was visible only at the beginning of the recordings, during baselines preceding Yoga intervention. This could be explained by the sympathetic activation through physical activity. Clark et al. investigated different autonomic measures for pregnant and non-pregnant women in various positions. 60 % of the pregnant women in the third trimester showed an unstable HR after the orthostatic maneuver to move from a supine to a standing position, which was reflected in a lower SampEn [44]. Even though we measured the entropy at the end of the Yoga lesson after a rest period, entropy may still be influenced by the previous physical activity. This suggests it may be more representative to measure Entropy Rate before the Yoga activity to assess a general stress reduction, and that the benefit of the Yoga activity cannot be captured by a higher Entropy Rate directly after the intervention.

The observed effect at the beginning of the recording was visible via the Entropy Rate but not the SampEn. This may be due to the Entropy Rate being a more robust biomarker of HRV complexity than SampEn under real-world recording conditions. Further studies are required to investigate this.

We did not detect any temporal, gestational changes in the control group, unlike some other studies, which showed changes due to the advancing pregnancy: Clark et al. showed that SampEn decreased from the first to the second and third trimesters [44]. Through continuous monitoring, Rowan et al. could observe a decrease in HRV in healthy pregnant women until 7 weeks before delivery [45]. Our last recording was at gestational week 35, equivalent to about 5 weeks before birth, and we observed this decrease in neither the control nor the Yoga group. It is possible that the natural progression of pregnancy would have led to altered HRVs in both groups, which were buffered in both groups as the women felt safer being observed within a study.

As early as the beginning of the study, both entropy biomarkers were already higher in the Yoga group than in the control group. The different entropy levels at study launch could be caused by a study bias due to different stress levels in both groups. However, the participants' perceived stress, measured by the PSS, was not significantly different at the study launch (see Table 1). It is conceivable that the terms of recording biased the results, even though we tried to make the conditions as favorable and similar as possible: we recorded at the same time of the day and suggested the supine position. However, when women felt discomfort, they were, of course, allowed to change to lying on one side, which could be a possible confounder.

The best-case explanation would be that the mere randomization into the Yoga group reduced stress. Knowing that the participants will receive a relaxation program and will be taken care of could impact the stress level positively, whereas some control group participants shared that the ECG appointments were burdensome for them.

#### 4.1.2. Comprehensive HRV analysis

To provide a more comprehensive perspective, we extended the analysis to 94 HRV metrics covering linear time, frequency, and nonlinear domains. By applying timepoint-specific principal component analysis (PCA), we derived a unified HRV index that captured dynamic variance structures across the four measurement periods. This approach confirmed that HRV metric relationships were not static but changed over time. Mixed effects models of the unified HRV index revealed a large and highly significant baseline difference between Yoga and control groups and a significant time × group interaction, indicating differential trajectories of HRV between groups across pregnancy rather than strictly parallel patterns. Together, these findings suggest that HRV patterns change with advancing pregnancy and are influenced by Yoga practice. These findings should be interpreted cautiously due to the small sample size.

*4.2. Relation to Previous Research*

To our knowledge, this is the first study to measure the effect of Yoga on chronically stressed pregnant women using entropy biomarkers. Few studies have previously assessed Yoga's effect on HRV in pregnant women.

Hayase et al. (2018) conducted a similar study to ours in Japan: pregnant women attending a Yoga class were compared to pregnant women who did not attend. Unlike in our study, the women were not screened to be stressed and were not randomized. HRV was assessed at different time points using a 24 h monitor, and results showed higher HRV in the Yoga group compared to controls at gestational week 28-31 at night and at gestational week 36-40 in the morning, afternoon, and late night [46]. This also indicates that it is challenging to grasp differences when analyzing at just one time point of the day, as in our case. Žebeljan et al. (2022) compared participants of a prenatal Yoga class with a control group, which practiced moderate-intensity walking, and observed higher HRV biomarkers in the Yoga group [47]. Such considerations highlight the potential value of wearables in providing continuous insights over longer periods of time and their ability to capture more subtle changes in HRV, which may yield novel biomarkers of health and disease.

As Yoga is an interface of physical activity, relaxation techniques, and spiritual aspects, it would be interesting to investigate further the difference between mere physical activity in pregnancy and Yoga practices. The literature shows inconsistent results about the impact of physical activity on maternal HRV. A systematic review by Dietz et al. (2016) observed some trends but did not prove the hypothesis of improved mHR due to physical activity, whereas a possible effect on fetal HRV was shown in most of the studies [48]. Malhotra et al. (2023) compared the effect of the practice of sun salutations for 20 minutes with 20 minutes of a mild intensity stationary bike session on healthy males. An influence on HRV was observed in the sun salutation group [49]. Satyapriya et al. (2009) conducted a randomized controlled trial in India. They compared a Yoga group, practicing Yoga and deep relaxation for 1 hour daily, with a control group that practiced standard prenatal exercises, including stretches and lying in the supine position. Results showed a stronger parasympathetic and lower sympathetic activity during the deep relaxation technique, which followed the Yoga exercises, as well as in the supine resting position, which followed the prenatal exercises, compared to measurements before the activity. The beneficial effect on the ANS was significant in the Yoga group, and the increase of the parasympathetic high-frequency band showed a 64% increase in pregnancy week 20, rising to 150% in pregnancy week 36, indicating that the continuous practice could improve the parasympathetic regulation. Remarkably, unlike in our study, the effects were not observed when comparing HRV from the beginning to the end of the study at baseline before the Yoga session, which in our case was the only time point at which we could observe an effect [42].

Interestingly, a systematic review found that Yoga trials conducted in India more often reported positive results [50]. However, the authors noted that publication bias and differences in trial reporting standards may contribute to this finding.

The entropy biomarkers were used successfully in other studies to evaluate stress increases or decreases. Byun et al. (2019) showed lower entropy features in patients with major depressive disorder than healthy ones; the difference was not significant during baseline recording but became more visible in the recovery phase after mental stress induction [20]. This may be why we could not show more pronounced group differences in our study, as we did not expose the participants to any controlled stressor. Furthermore, in the mentioned study, the entropy values were still recovering 15 minutes after the stressor, indicating prolonged sympathetic activity. This raises the question of whether, after movement in the Yoga class, we captured enduring sympathetic excitation in the measurements at the end, which might disguise a possible positive effect of the Yoga intervention.

*4.3. Strengths and Study Limitations*

4.3.1. Strengths

Our study design gives us an insight into the ANS during pregnancy with and without intervention: We assessed the women longitudinally for around 20 weeks, covering the second and the third trimesters, instead of observing them cross-sectionally at one time point in pregnancy. Screening the women for chronic stress helped make the effect of a Yoga intervention clearer. The entropy biomarkers gave us an insight into the ANS. By using the self-reported stress questionnaire, PSS, and the biomarkers, we assessed stress from two different angles.

We chose the most commonly used biomarker of HRV complexity, SampEn, and its more cutting-edge counterpart, Entropy Rate, which is known to be more robust and also theoretically grounded for characterizing signal complexity. The Entropy Rate captures better temporal dependencies by measuring the average uncertainty (or new information) per symbol, taking into account the conditional dependencies between the sequential data points. This means it reflects how predictable a signal is over time. In contrast, SampEn looks at the similarity of fixed-length patterns in the data without fully capturing long-range temporal dependencies. Entropy Rate is rooted in information theory (related to concepts like Kolmogorov–Sinai Entropy) and represents the asymptotic limit of the per-symbol uncertainty of a stochastic process. This gives it a strong theoretical backing when it comes to describing the inherent complexity of a process. Conversely, SampEn requires careful selection of parameters such as the embedding dimension and tolerance level. Its estimates can be sensitive to these choices, especially for shorter or noisy time series. Entropy Rate, being an asymptotic measure, is less sensitive to these finite-sample issues when sufficient data is available. While SampEn provides an estimate based on observed pattern matches in finite data, it may not converge to the actual underlying complexity if the data set is short or the patterns are not well-represented. Entropy Rate, on the other hand, is defined as the limit of these finite-sample measures as the sample size grows, offering a more accurate reflection of the signal's complexity in the long run. In summary, while SampEn is useful for practical applications with limited data, the Entropy Rate gives a more complete and less parameter-dependent picture of signal complexity by fully accounting for the dynamics and dependencies present in the signal.

We characterized HRV from the following two complementary angles.

After establishing SampEn and Entropy Rate as complementary indicators of HRV complexity, we broadened the scope to include a total of 94 HRV metrics from time-, frequency-, and nonlinear domains. This expansion allowed us to capture a much wider spectrum of autonomic nervous system dynamics and to detect patterns that might remain hidden when relying on a narrow subset of HRV biomarkers.

This breadth was complemented by methodological advances that strengthened the robustness of the findings. By adjusting for maternal age, BMI, and gestational age, we reduced the influence of key confounders. Timepoint-specific PCA further enabled us to account for changing interrelationships among HRV metrics across pregnancy, while the resulting unified HRV index condensed this multidimensional information into a single interpretable score. In addition, the use of linear mixed-effects models provided a rigorous framework for handling repeated measures and testing for intervention effects, even within the constraints of a modest sample size and identified baseline group differences.

A further strength lies in our use of a validated, automated HRV analysis pipeline [41]. This approach enhances reproducibility and transparency by making the code publicly accessible, reducing subjective analytic choices, and ensuring scalability for larger datasets and future studies. By minimizing manual intervention, the pipeline establishes a robust and standardized framework for multidimensional HRV analysis.

### 4.3.2. Limitations

Our study has some limitations that should be noted. First, our study's sample size is relatively small, and due to a low number of participant requests, we employed quasi-randomization rather than the planned full randomization. This approach may introduce allocation bias, thereby limiting the internal validity and generalizability of our findings. Furthermore, the reduced sample size decreased the statistical power from the aimed 80% to 56%, increasing the likelihood of type II error.

In addition, entropy values could not be extracted for some measurements, which further reduced the amount of usable data. This limitation required adjustments of the analysis window for certain sessions, as described above, and in some cases prevented the use of recordings from the very beginning or end of the study. For some participants, there was an accumulation of non-evaluable recordings, which raised the question whether individual factors (for example, the skin structure or anatomical differences) might have limited the device functionality. For the delta analyses, the number of available data points was very limited, which should be taken into account when interpreting the results. Furthermore, the unusable data led to unbalanced sample sizes (Yoga n=6-8 vs control n=12 per timepoint) and might limit the generalizability of the results

Second, the entropy baseline at the study begin was already higher in the Yoga group, which introduces concerns with the validity of the analyses on absolute differences, but is handled by the delta analyses.

Third, the questionnaires we used for screening in this study were specifically designed to assess self-perceived chronic stress and anxiety. We did not explicitly assess depressive symptoms, although there is a considerable symptom overlap between anxiety, stress, and depression, which introduces the risk of unmeasured depressive symptoms as a possible confounder.

Fourth, not all participants in the intervention group attended the required 75% of the Yoga or Yoga Nidra sessions, and some participants in the control group also practiced Yoga independently. As all study applicants were naturally interested in Yoga, restricting their practice would have been ethically problematic. Nevertheless, the intervention group engaged in significantly more minutes of Yoga per week, allowing for a comparison of the effects between the groups. In addition, we did not collect detailed information on participants' physical activity history prior to pregnancy. This limits our ability to fully account for potential pre-existing lifestyle differences that may have influenced both intervention uptake and physiological response. While our expanded statistical framework adjusted for key covariates such as maternal age, gestational age, and BMI, the absence of pre-intervention activity data remains an important limitation that should be considered when interpreting the findings.

A further consideration is that the duration and frequency of the ECG recordings differ between groups, introducing a measurement bias.

In addition, the study began in the second trimester of pregnancy, meaning we could not capture potential changes in the first trimester.

It should also be noted that the effects of Hatha Yoga, Yoga Nidra, and general physical activity cannot be separated in this study, and all components may have partly contributed to stress reduction, representing a potential source of confounding.

Finally, the study sample consisted predominantly of women with a high level of education and income, which may limit the generalizability of the results to the broader population of pregnant women experiencing chronic stress.

Future studies should aim to address these limitations to validate and extend our findings.

*4.4. Implications and Future Research*

*4.4.1. Wearables and Feedback*

Yuan et al. (2019) designed a portable ECG device that can send data to a smartphone. The device recorded mECG, and an algorithm detected and extracted the fHR. Tests suggested that the technique could work as a continuous monitoring to show pathologies and enable early medical intervention when necessary [51]. Continuous noninvasive monitoring could hence be used as a feedback tool for pregnant women, which shows the mothers when they and/ or the child are under stress, and they could benefit even more from a relaxation program like Yoga.

Jasinski et al. (2024) conducted a retrospective case-control study on data obtained by wearables that continuously monitored HRV and resting HR in pregnancy. Preterm births were compared to term births, and results suggest that an inflection in nightly maternal HRV is an indicator of time until birth for term as well as preterm pregnancies. This indicates that continuous HRV monitoring could serve as a screening tool for preterm birth to alert for further investigations [52].

Future investigations should prioritize larger, balanced samples with randomized controlled designs to confirm causal relationships. Extending follow-up periods could help determine whether the observed trajectory patterns continue throughout pregnancy and into the postpartum period. Further, examining dose–response relationships may inform optimal practice prescriptions for clinical implementation. Building on our analytical framework, future studies should also incorporate additional physiological signals and leverage advanced computational methods, such as sensor fusion and machine learning, to more precisely characterize autonomic regulation and intervention effects in larger and more diverse cohorts.

*4.4.2. Relation of Maternal Entropy and Fetal Health*

In the previous study, Lobmaier et al. (2020) observed a coupling of mHRs and fHRs, showing that for chronically stressed pregnant women, the fHR more likely reacted to drops of mHR. In contrast, in the non-stressed group, the fHR remained more stable, i.e., less responsive to mHR decelerations [15].

In this study, we focused on the mHR characteristics in response to chronic stress because these are more ubiquitously accessible than fHR. Nonetheless, we hypothesize that in this chronically stressed cohort, the mHR dynamics transfer to the fHR dynamics, as shown in the previous study. As such, our insights into the effects of Yoga on mHR may indirectly point to a potential impact on the fHR as well. These interpretations remain speculative, given that no fetal measurements were obtained in this study. However, AI-based analyses of our dataset may in the future allow extraction of fetal HR characteristics and enable us to directly test whether Yoga positively affects fetal physiology. Future studies, ideally combining maternal and fetal HR monitoring, are needed to clarify potential causal pathways. Finding methods to lower PS is essential to protect fetuses from adverse effects on their development.

# 5. Conclusion

In this study, we examined maternal HRV complexity in the context of a prenatal Yoga intervention. Initial analyses of SampEn and Entropy Rate suggested group differences, but these effects disappeared after adjusting for baseline imbalances and relevant covariates, indicating that the apparent effects were primarily driven by allocation bias rather than the intervention itself.

By extending the analysis to a comprehensive set of 94 HRV metrics with covariate adjustment and timepoint-specific PCA, we derived a unified HRV index that captured dynamic variance structures across pregnancy. Across both groups, correlation analyses revealed a simplification of the HRV correlation network over time. Importantly, the PCA framework allowed us to determine which physiological domains contributed most strongly to overall HRV variability at each measurement point, providing deeper insight into the evolving architecture of autonomic complexity.

We could demonstrate a restructuring of domain contributions, with frequency measures becoming less prominent and complexity measures gaining importance in late pregnancy. These general observations suggest that pregnancy itself is associated with dynamic shifts in the structure of autonomic regulation. Mixed-effects models applied to the unified index detected significant baseline differences and a significant time × group interaction, suggesting differential trajectories of HRV between Yoga and control groups across pregnancy.

Taken together, these findings underscore both the challenges and the opportunities in studying maternal autonomic regulation during pregnancy. While causal interpretation remains limited by baseline imbalances and small sample size, the results demonstrate the feasibility and added value of advanced HRV analysis frameworks to capture dynamic, multidimensional patterns. This methodological groundwork provides a foundation for future, larger studies to disentangle true intervention effects from baseline differences and to establish robust biomarkers of mind–body interventions in pregnancy.

**Author Contributions:** Conceptualization, Marlene J. E. Mayer, Clara Becker, Marta C. Antonelli, Silvia M. Lobmaier, and Martin G. Frasch; methodology, Marlene J. E. Mayer, Clara Becker, Marta C. Antonelli, Silvia M. Lobmaier, and Martin G. Frasch; software, Nicolas B. Garnier and Martin G. Frasch; validation, Silvia M. Lobmaier and Martin G. Frasch; formal analysis, Marlene J. E. Mayer, Nicolas B. Garnier, and Martin G. Frasch; investigation, Marlene J. E. Mayer, Clara Becker, Marta C. Antonelli, and Silvia M. Lobmaier; resources, Marta C. Antonelli, Silvia M. Lobmaier, and Martin G. Frasch; data curation, Marlene J. E. Mayer and Clara Becker; writing—original draft preparation, Marlene J. E. Mayer; writing—review and editing, Clara Becker, Marta C. Antonelli, Silvia M. Lobmaier, and Martin G. Frasch; visualization, Marlene J. E. Mayer and Martin G. Frasch; supervision, Marta C. Antonelli, Silvia M. Lobmaier, and Martin G. Frasch; project administration, Marlene J. E. Mayer, Clara Becker, Marta C. Antonelli, and Silvia M. Lobmaier; funding acquisition, Silvia M. Lobmaier. All authors have read and agreed to the published version of the manuscript.

**Funding:** This research did not receive any specific grant from agencies in the public, commercial, or not-for-profit sectors. We used partial funding from the Institute of Advanced Studies of the Technical University of Munich and the Dr. Geisenhofer Foundation (Munich, Germany). Bittium Corporation subsidized the acquisition of the Faros 360 ECG devices used in this project.

**Institutional Review Board Statement:** The study was conducted in accordance with the Declaration of Helsinki and approved by the Ethics Committee at the TUM University Hospital under registration number 2022-86-S-SR on 6 September 2022. The study was retrospectively registered as a clinical trial on 24 July 2025 with the registration number: DRKS00037529.

**Informed Consent Statement:** Informed consent was obtained from all subjects involved in the study.

**Data Availability Statement:** The datasets presented in this article are not yet publicly available, as they are intended to be used in a forthcoming publication.

**Acknowledgments:** We would like to thank all participants for attending our study. We greatly thank Anne Loewer, our experienced prenatal Yoga teacher, for all her efforts in designing and implementing the Yoga lessons. We thank PD Dr. F. Stumpfe for the support in recruiting participants from the "Klinikum Dritter Orden" for the study. During the preparation of this work, the authors used Chat-GPT and Grammarly for the purpose of phrasing support and proofreading. The authors have reviewed and edited the output and take full responsibility for the content of this publication.

**Conflicts of Interest:** Silvia M. Lobmaier reports financial support provided by the Institute of Advanced Studies of the Technical University of Munich and the Dr. Geisenhofer Foundation (Munich, Germany), as well as equipment provided by Bittium. Martin G. Frasch reports relationships with several organizations, including board membership, equity or stock ownership, and travel reimbursement from Nurtur Health Inc.; board membership and equity or stock ownership in JoyBucket Corporation and Health Stream Analytics LLC; and consulting, equity or stock ownership, and funding grants from Delfina Care Inc. He also reports pending patents with the University of Washington and the University of Toronto, as well as with Health Stream Analytics LLC. Martin G. Frasch serves as a Guest Editor for the *Bioengineering* special issue *"Fetal-Maternal Monitoring during Pregnancy and Labor:*

*Trends and Opportunities."* All other authors declare no known competing financial interests or personal relationships that could have appeared to influence the work reported in this paper.

## Abbreviations

The following abbreviations are used in this manuscript:

| | |
|---|---|
| PS | Prenatal stress |
| HRV | Heart rate variability |
| ANS | Autonomic nervous system |
| HPA axis | Hypothalamic-Pituitary-Adrenal axis |
| aECG | Transabdominal electrocardiogram |
| mHR | Maternal heart rate |
| fHR | Fetal heart rate |
| FSI | Fetal stress index |
| PCA | Principal Component Analysis |
| SampEn | Sample Entropy |
| PC | Principal component |

# Supplementary material

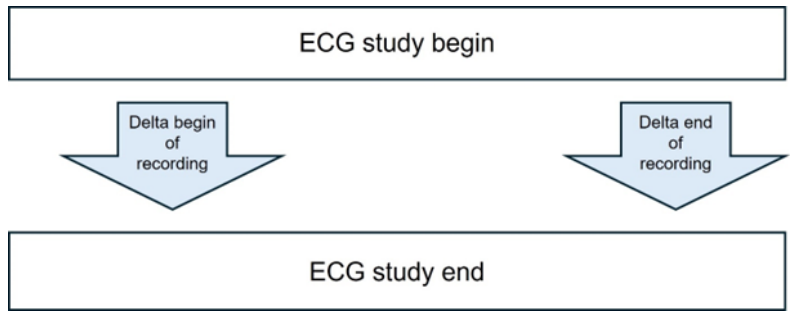

**Figure S1.** Delta analyses show the entropy differences from the beginning to the end of the study, with the aim of showing Yoga intervention-related changes. We assessed the Delta values both at baseline before the Yoga lesson/ at the beginning of the recording as well as after the Yoga lesson/ at the end of the recording.

**Table S1. 94 HRV metrics used for analysis**

| Rank | Metric Name | Category | Description |
|---|---|---|---|
| 1 | MeanNN | Time Domain | Time domain HRV metric: MeanNN |
| 2 | SDNN | Time Domain | Standard Deviation of Normal-to-Normal intervals |
| 3 | SDANN1 | Time Domain | SDANN - standard deviation of average NN in segments |
| 4 | SDNNI1 | Time Domain | SDNNI - mean of standard deviations in segments |
| 5 | SDANN2 | Time Domain | SDANN - standard deviation of average NN in segments |
| 6 | SDNNI2 | Time Domain | SDNNI - mean of standard deviations in segments |
| 7 | SDANN5 | Time Domain | SDANN - standard deviation of average NN in segments |
| 8 | SDNNI5 | Time Domain | SDNNI - mean of standard deviations in segments |
| 9 | RMSSD | Time Domain | Root Mean Square of Successive Differences |
| 10 | SDSD | Time Domain | Time domain HRV metric: SDSD |
| 11 | CVNN | Time Domain | Coefficient of variation measure |
| 12 | CVSD | Time Domain | Coefficient of variation measure |
| 13 | MedianNN | Time Domain | Time domain HRV metric: MedianNN |
| 14 | MadNN | Time Domain | Time domain HRV metric: MadNN |
| 15 | MCVNN | Time Domain | Time domain HRV metric: MCVNN |
| 16 | IQRNN | Time Domain | Time domain HRV metric: IQRNN |
| 17 | SDRMSSD | Time Domain | Time domain HRV metric: SDRMSSD |
| 18 | Prc20NN | Time Domain | Time domain HRV metric: Prc20NN |
| 19 | Prc80NN | Time Domain | Time domain HRV metric: Prc80NN |
| 20 | pNN50 | Time Domain | Percentage of NN intervals > 50ms different |
| 21 | pNN20 | Time Domain | Percentage of NN intervals > 20ms different |
| 22 | MinNN | Time Domain | Time domain HRV metric: MinNN |
| 23 | MaxNN | Time Domain | Time domain HRV metric: MaxNN |
| 24 | HTI | Time Domain | Time domain HRV metric: HTI |
| 25 | TINN | Time Domain | Triangular Interpolation of NN interval histogram |
| 26 | LF | Frequency Domain | Low Frequency power (0.04-0.15 Hz) |
| 27 | HF | Frequency Domain | High Frequency power (0.15-0.4 Hz) |
| 28 | TP | Frequency Domain | Total power |
| 29 | LF_HF | Frequency Domain | LF/HF ratio |
| 30 | LFnu | Frequency Domain | Low Frequency power normalized (0.04-0.15 Hz) |
| 31 | HFnu | Frequency Domain | High Frequency power normalized (0.15-0.4 Hz) |
| 32 | SD1 | Nonlinear | Poincaré plot geometric measure: SD1 |
| 33 | SD2 | Nonlinear | Poincaré plot geometric measure: SD2 |
| 34 | SD1SD2 | Nonlinear | Poincaré plot geometric measure: SD1SD2 |
| 35 | S | Nonlinear | Nonlinear HRV metric: S |
| 36 | CSI | Nonlinear | Nonlinear HRV metric: CSI |
| 37 | CVI | Nonlinear | Nonlinear HRV metric: CVI |
| 38 | CSI_Modified | Nonlinear | Nonlinear HRV metric: CSI_Modified |
| 39 | PIP | Nonlinear | Nonlinear HRV metric: PIP |
| 40 | IALS | Nonlinear | Nonlinear HRV metric: IALS |
| 41 | PSS | Nonlinear | Nonlinear HRV metric: PSS |
| 42 | PAS | Nonlinear | Nonlinear HRV metric: PAS |
| 43 | GI | Nonlinear | Nonlinear HRV metric: GI |
| 44 | SI | Nonlinear | Nonlinear HRV metric: SI |
| 45 | AI | Nonlinear | Nonlinear HRV metric: AI |
| 46 | PI | Nonlinear | Nonlinear HRV metric: PI |
| 47 | C1d | Nonlinear | Nonlinear HRV metric: C1d |
| 48 | C1a | Nonlinear | Nonlinear HRV metric: C1a |
| 49 | SD1d | Nonlinear | Poincaré plot geometric measure: SD1d |
| 50 | SD1a | Nonlinear | Poincaré plot geometric measure: SD1a |
| 51 | C2d | Nonlinear | Nonlinear HRV metric: C2d |
| 52 | C2a | Nonlinear | Nonlinear HRV metric: C2a |
| 53 | SD2d | Nonlinear | Poincaré plot geometric measure: SD2d |
| 54 | SD2a | Nonlinear | Poincaré plot geometric measure: SD2a |
| 55 | Cd | Nonlinear | Nonlinear HRV metric: Cd |
| 56 | Ca | Nonlinear | Nonlinear HRV metric: Ca |
| 57 | SDNNd | Nonlinear | Poincaré plot geometric measure: SDNNd |
| 58 | SDNNa | Nonlinear | Poincaré plot geometric measure: SDNNa |
| 59 | DFA_alpha1 | Nonlinear | Nonlinear HRV metric: DFA_alpha1 |
| 60 | MFDFA_alpha1_Width | Nonlinear | Multifractal Detrended Fluctuation Analysis parameter |
| 61 | MFDFA_alpha1_Peak | Nonlinear | Multifractal Detrended Fluctuation Analysis parameter |
| 62 | MFDFA_alpha1_Mean | Nonlinear | Multifractal Detrended Fluctuation Analysis parameter |
| 63 | MFDFA_alpha1_Max | Nonlinear | Multifractal Detrended Fluctuation Analysis parameter |
| 64 | MFDFA_alpha1_Delta | Nonlinear | Multifractal Detrended Fluctuation Analysis parameter |
| 65 | MFDFA_alpha1_Asymmetry | Nonlinear | Multifractal Detrended Fluctuation Analysis parameter |
| 66 | MFDFA_alpha1_Fluctuation | Nonlinear | Multifractal Detrended Fluctuation Analysis parameter |
| 67 | MFDFA_alpha1_Increment | Nonlinear | Multifractal Detrended Fluctuation Analysis parameter |
| 68 | DFA_alpha2 | Nonlinear | Nonlinear HRV metric: DFA_alpha2 |
| 69 | MFDFA_alpha2_Width | Nonlinear | Multifractal Detrended Fluctuation Analysis parameter |
| 70 | MFDFA_alpha2_Peak | Nonlinear | Multifractal Detrended Fluctuation Analysis parameter |
| 71 | MFDFA_alpha2_Mean | Nonlinear | Multifractal Detrended Fluctuation Analysis parameter |
| 72 | MFDFA_alpha2_Max | Nonlinear | Multifractal Detrended Fluctuation Analysis parameter |
| 73 | MFDFA_alpha2_Delta | Nonlinear | Multifractal Detrended Fluctuation Analysis parameter |
| 74 | MFDFA_alpha2_Asymmetry | Nonlinear | Multifractal Detrended Fluctuation Analysis parameter |
| 75 | MFDFA_alpha2_Fluctuation | Nonlinear | Multifractal Detrended Fluctuation Analysis parameter |
| 76 | MFDFA_alpha2_Increment | Nonlinear | Multifractal Detrended Fluctuation Analysis parameter |
| 77 | ApEn | Nonlinear | Approximate Entropy - complexity measure |
| 78 | SampEn | Nonlinear | Sample Entropy - regularity measure |
| 79 | ShanEn | Nonlinear | Entropy measure: ShanEn |
| 80 | FuzzyEn | Nonlinear | Entropy measure: FuzzyEn |
| 81 | MSEn | Nonlinear | Entropy measure: MSEn |
| 82 | CMSEn | Nonlinear | Entropy measure: CMSEn |
| 83 | RCMSEn | Nonlinear | Entropy measure: RCMSEn |
| 84 | CD | Nonlinear | Nonlinear HRV metric: CD |
| 85 | HFD | Nonlinear | Nonlinear HRV metric: HFD |
| 86 | KFD | Nonlinear | Nonlinear HRV metric: KFD |
| 87 | LZC | Nonlinear | Nonlinear HRV metric: LZC |
| 88 | EntropyRate | Nonlinear | Novel entropy rate measure - our methodological contribution |
| 89 | additional_hrt_turbulence_onset | Other | HRV metric: additional_hrt_turbulence_onset |
| 90 | additional_hrt_turbulence_slope | Other | HRV metric: additional_hrt_turbulence_slope |
| 91 | additional_coefficient_variation | Other | HRV metric: additional_coefficient_variation |
| 92 | additional_temporal_variability | Other | HRV metric: additional_temporal_variability |
| 93 | additional_spectral_centroid | Other | HRV metric: additional_spectral_centroid |
| 94 | additional_spectral_bandwidth | Other | HRV metric: additional_spectral_bandwidth |